\documentclass[useAMS,usenatbib,usedcolumn]{mn2e}

\usepackage{graphicx}

\def\cf{{cf.~}}
\def\ie{{i.e.,~}}
\def\eg{{e.g.,~}}
\def\hinv{$h^{-1}$}
\def\kms{~{\rm km~s^{-1}}}

\newcommand{\ig}{inter-galactic }
\newcommand{\igm}{inter-galactic medium }
\newcommand{\gr}{$\gamma$-ray }
\newcommand{\pnd}{$\pi^0$-decay }
\newcommand{\lowa}{Loeb \& Waxman \/}
\newcommand{\epm}{e$^\pm$ }

\newcommand{\ic}{IC }

\begin{document}
\title[IG Shock Acceleration and the CGB]
  {Inter-galactic Shock Acceleration and the Cosmic Gamma-ray Background}
\author[F.\ Miniati]
  {Francesco~Miniati\thanks{fm@MPA-Garching.MPG.DE}\\
  Max-Planck-Institut f\"ur Astrophysik,
     Karl-Schwarzschild-Str. 1, 85740, Garching, Germany}
\date{\today}
\pubyear{2001} \volume{000} \pagerange{1} \onecolumn

\maketitle \label{firstpage}

\begin{abstract}
We investigate {\it numerically} the contribution
to the cosmic gamma-ray background from cosmic-rays
ions and electrons accelerated
at \ig shocks associated with cosmological structure formation.
We show that the kinetic energy of accretion flows
in the low-red-shift inter-galactic medium is thermalized primarily
through moderately strong shocks, which allow for an efficient
conversion of shock ram pressure into cosmic-ray pressure.
Cosmic-rays accelerated at these shocks produce a diffuse gamma-ray flux 
which is dominated by inverse Compton emission from 
electrons scattering off cosmic microwave background photons. 
Decay of neutral $\pi$-mesons generated in p-p inelastic collisions
of the ionic cosmic-ray component with the thermal gas
contribute about 30 \% of the computed emission.
Based on experimental upper limits on the photon flux above 
100 MeV from nearby clusters we
constrain the efficiency of conversion of shock ram pressure
into relativistic CR electrons to $\la 1\%$.
Thus, we find that cosmic-rays of cosmological
origin can generate an overall significant fraction of 
order 20 \% and no more than 30 \% 
of the measured gamma-ray background.

\end{abstract}

\begin{keywords}
acceleration of particles  ---  cosmology: large-scale structure
of universe  ---  galaxies: clusters: general  --- 
gamma rays: theory  ---  methods: numerical  --- 
radiation mechanism: non-thermal  ---   shock waves 
\end{keywords}

\section{Introduction}

After the re-ionization epoch, the inter-galactic medium is heated up to 
temperatures
$T\geq 10^4$ K by energy input from quasars and star forming regions.
As the wavelength of primordial density perturbations
entering the non-linear stage, $\lambda_{NL}(z)$, grows longer in size,
the temperature of the low red-shift gas is further raised by shock 
waves.
These convert the specific kinetic energy associated with accretion flows,
of order $\lambda_{NL}^2(z) H^2(z)$ - with $H(z)$ the Hubble parameter 
and $z$ the cosmological red-shift -
into thermal energy corresponding to 7-8 keV per baryon in massive
clusters of galaxies and below $\sim 1$ keV in the diffuse inter-galactic 
medium \citep{ceos99a}.
The importance of this process in cosmic history is
emphasized by the fact that on the one hand
the large scale structure is most prominent in
thermal X-ray radiation by hot intra-cluster gas. On the other hand
numerical simulations predict that most of the baryons,
which are observed at high red-shift as Ly-$\alpha$ absorbers,
end up shock-heated to temperatures in the range $10^5-10^7$ K
in the contemporary universe \citep{ceos99a}.

Shock waves in cosmic environment are collision-less meaning that
the dissipation process occurs via random electro-magnetic fields
in the turbulent post-shock flow. In addition to a thermal population of
hot gas, they generate a supra-thermal distribution of
high energy particles, generally referred to as cosmic-rays (CR).
Shock acceleration of CR electrons up to TeV energies is directly 
observed in supernova remnants 
\citep{allengetal97,koyamaetal97,tanimorietal98,muraishietal00,alpego01}
and the bulk of the Galactic CR ion population up to
$10^{14}-10^{15}$ eV is also thought to originate there. The same
acceleration mechanism is expected at work at large scale structure 
shocks.
Indeed, we do observe non-thermal radiation in several massive
clusters of galaxies both at radio \citep[\eg][and references therein]{feretti99}
and hard X-ray frequencies \citep[][and references therein]{fufeetal99,fufeetal00}.
At least in some models CR shock acceleration of both ions and electrons
seems an essential ingredient to explain different features of the 
observed radio emission \citep[\eg][]{mjkr01}. 
In addition, the possibility that CRs be accelerated at 
\ig shocks has raised the issue that CR ions, due to their
long lifetime against energy losses and to their efficient 
confinement by magnetic irregularities \citep{voahbr96,cobl98}, 
may accumulate inside forming structures and store a 
significant fraction of the total pressure
there \citep{mint00,mrkj01}.
Beside cosmic shocks, the energetics characterizing other CR sources
such as termination shocks driven by galactic winds
\citep{voahbr96} and radio galaxies \citep{ebkw97} 
are also noteworthy.

Along these lines of observational and theoretical progress, 
the idea has recently been put forth that CR electrons accelerated
at \ig shocks may generate the observed \gr
background (CGB) by way of \ic (IC) scattering off
cosmic microwave background (CMB) photons \citep{lowa00}.
The CGB was measured by EGRET onboard the {\it Compton Gamma Ray
Observatory}. It consists of a differential flux, 
$\varepsilon^2 J_\varepsilon \sim $
5.4 ($\varepsilon$ /keV)$^{-0.1}$ keV cm$^{-2}$ s$^{-1}$ sr$^{-1}$ \citep{sreeku98}
throughout an energy range extending from about 100 MeV to 100 GeV.
The integrated photon flux above 100 MeV
corresponds to 1.45 $\times 10^{-5}$ ph cm$^{-2}$ s$^{-1}$ sr$^{-1}$
\citep{sreeku98}, in agreement with previously measures 
carried out by the {\it SAS 2} satellite \citep{thfi82}.
Its origin is most intriguing. 
The contribution from identified EGRET blazars to the integrated
photon flux above 100 MeV amounts
to 1$\times 10^{-6}$ ph cm$^{-2}$ s$^{-1}$ sr$^{-1}$ \citep{much99};
that is $\sim$ 7 \% of the measured CGB flux in corresponding units. 
A number of authors have suggested that the contribution from an
unresolved, faint end of the blazar $\gamma$-ray distribution function 
could fill the gap between the observed flux and that which is already 
accounted for.
And in fact, though with large uncertainties, some authors were 
able to validate the above conjecture 
by using a \gr luminosity function derived from the 
blazar radio luminosity function and the assumption of a 
constant ratio of radio to \gr fluxes 
\citep[\eg][]{pgfc93,stsa96}.
However, the correlation between the blazar \gr and radio flux is
only weakly established and the existing evidence might indeed be
dominated by bias effects \citep{mueckeetal97}.
In fact, based solely on the observationally derived
luminosity function of \gr loud AGNs that have been detected by the 
EGRET experiment, \citet{chmu98} found that only 
25 \% of the CGB can be accounted for by undetected \gr loud
blazars. 
Alternatively, the idea was put forward that the CGB is generated 
by the decay of neutral $\pi$-mesons, $\pi^0 \rightarrow 
\gamma \gamma$, generated by an all pervasive population of CR ions 
interacting via p-p collisions with the nuclei of the diffuse 
\igm \citep{dasha95} . However, it was later argued that the CR ions 
from the available sources of CRs in the IGM, namely shocks, 
Active Galactic Nuclei and normal galaxies, could provide at most
a fraction of order of a \% of the observed CGB \citep{bbp97,cobl98}. 

In this paper we carry out a numerical study to investigate 
the contribution to the CGB from CRs accelerated at cosmological, 
\ig shocks associated with structure formation. We show that 
the heating of the IGM is due primarily to moderately strong 
shocks, which allow for some fraction of the 
shock ram pressure to be converted into CR pressure.
According to our results, the contribution to the CGB from 
cosmological CRs could indeed be significant. It is 
dominated by IC emission from primary 
electrons which, however, we estimate to be smaller than previously 
found \citep{lowa00}. In addition we find a \gr 
flux from $\pi^0$-decay that is not negligible and larger than, 
although of the same order as, the upper limits estimated by \cite{cobl98}.

The paper is organized as follows:
we describe the numerical model in \S \ref{numsim.se},
present the results in \S \ref{res.se} discuss them in 
\S \ref{disc.se} and summarize with \S \ref{con.se}.

\section{The Simulation} \label{numsim.se}

Our investigation is based on numerical
simulations of large scale structure formation which include
shock acceleration, transport and energy losses/gains 
of CR ions and electrons.
In the following sections we describe the salient 
features of the employed numerical techniques.

\subsection{Cosmological Model}

The formation and evolution of the 
large scale structure is computed by means of an
Eulerian, grid based Total-Variation-Diminishing
hydro+N-body code \citep{rokc93}. The code is capable of 
sharply capturing both very strong and weak shocks while being
computationally relatively inexpensive. This feature is of primary 
importance in this kind of study since cosmic shocks have 
direct impact on the CR properties. 

We focus on a single, currently favored $\Lambda$CDM 
``concordance model'' \citep{osst95}. 
The assumed model is flat \citep[\eg][]{jaffeetal01}
with a total mass density $\Omega_m=0.3$ and a
vacuum energy density $\Omega_\Lambda= 1- \Omega_m= 0.7$.
The normalized Hubble constant is taken to be 
$h\equiv H_0/100$ km s$^{-1}$ Mpc$^{-1}$ = 0.67 \citep{freedman00}
and the baryonic mass density, 
$\Omega_b=0.04$. This value, after Big Bang nucleosynthesis 
calculations \citep[\eg][]{olstwa00} 
is only marginally consistent (at the 2 $\sigma$ level)
with the best fit to cosmic microwave background (CMB) 
measurements \citep[\eg][]{jaffeetal01}.
The initial density
perturbations are generated as a Gaussian random field with a power spectrum
characterized by a spectral index $n_s=1$ and ``cluster-normalization'' 
$\sigma_8=0.9$. The initial velocity field is computed through 
the Zel'dovich approximation.
The computational box is set to a comoving size $L=50$ \hinv Mpc.
The dark matter component is described by 256$^3$ particles
whereas the gas component is evolved on a comoving grid of 512$^3$ 
zones. Thus each numerical cell measures about 100 \hinv kpc
(comoving) and each dark 
matter particle corresponds to $2\times 10^9$ \hinv M$_\odot$.

\subsection{Cosmic-rays} \label{cosray.se}

The CR dynamics is computed numerically through the code COSMOCR 
\citep{min01}. 
The code follows simultaneously three CR populations, namely 
primary ions and electrons and secondary electrons/positrons ($e^\pm$). 
Primary ions and electrons are injected and accelerated 
at cosmic shocks according to the diffusive shock acceleration 
(DSA) mechanism \cite[\eg][]{blei87}. 
Secondary \epm are produced 
in p-p inelastic collisions of CR ions with thermal IGM nuclei.
In addition the code accounts for spatial transport and energy
losses/gains undergone by each CR component. 
Before describing the numerical technique in some detail 
(\S \ref{cosrayion.se} and \S \ref{cosrayele.se}) we address 
some of the physical assumptions underlying our model.

In general CRs are treated as passive quantities, meaning that at this stage 
their dynamical role is completely neglected both on the shock structure 
({\it test particle limit}) and the gas dynamics.
Recently developed kinetic models of DSA theory
certainly allow for solutions in which shock acceleration is highly 
efficient and the CRs strongly affect the shock structure \citep[\eg][]{madr01}. 
In these solutions, however, 
most of the shock associated ram pressure is converted into CR pressure
which becomes the dominant dynamic component. Although it cannot be 
completely ruled out that such a situation occurs in the cosmic environment,
both dynamical considerations 
\citep{mavi97,hms99,nemafo00,rosabl00,miba95,wu00}
as well as existing upper limits on \gr emission from nearby
clusters \citep{blasi99,mrkj01}, suggest that the ICM pressure is
dominated by a thermal component, with room for a still significant
CR pressure component at the level of a few tens of percent.
The latter is likely to affect the gas-dynamics of the intra-cluster (and
intergalactic medium) and will be considered in future work.  
  
Also, we implicitly assume the existence of a diffuse background 
magnetic field which is essential not only for the acceleration 
of the CRs but also for their subsequent confinement 
within cosmic structures. In support of this assumption we
notice that inter-galactic magnetic fields are indeed observed 
in the media of clusters of galaxies at the 
$\mu$G level \citep{clkrbo00}. Observations are much more difficult 
for the case of filaments and therefore the evidence there is
more sparse. However, the cases of rotation measure from the 
Coma super-clusters \citep{kkgv89} as well as the recent detection
of a radio filament \citep{bagchietal02} suggest that indeed 
sub-$\mu$G magnetic fields might be common in the IGM
\citep{kronberg94}.

\subsubsection{Cosmic-ray Ions} \label{cosrayion.se}

The first step in simulating CRs consist in detecting shocks where 
their injection and acceleration takes place.
Shocks are identified when a converging flow
(${\bf \nabla \cdot v < 0}$) experiences a pressure
jump $\Delta P/P$ above a given threshold.
For the present simulation, such threshold was set to the 
pressure jump across a shock with Mach number M=1.5.
Successful detection of shocks requires a scheme that reproduces 
them as sharp transition in the numerical solution. 
This is especially important 
because relatively weak shocks, which tend to be harder to capture, 
cannot be neglected. In fact, they turn out to process a substantial
fraction (but not all) of the kinetic energy associated to cosmic 
flows \citep[][ \S \ref{igms.se}]{minetal00}. 

The simulated CR ions are supposed to be injected in the DSA  
mechanism from the thermal IGM. Here the injection process is modeled 
after the thermal leakage prescription \citep[\eg][]{kajo95}.
We assume that, upon shock passage, the gas thermalizes to a Maxwellian 
distribution defined by a post-shock temperature $T_{\rm shock}$. 
Thermal ions in the high energy end of such a distribution might be fast 
enough to ``leak'' back upstream of the shock against trapping by 
the plasma waves that moderate the shock itself 
\citep{elei84,kajo95,mavo95}.
These ions are thus injected from the thermal pool in the DSA 
mechanism and quickly gain energy much above the thermal average 
\citep[\eg][]{blei87}.
As in \citet{mrkj01}, based on cluster/group properties,
we it found appropriate to follow  CR ions up to momenta
$p_{max}= 10^6$GeV/c. In fact,
CR ions up to energies of $10^{9}-10^{10}$ GeV/c can be accelerated at
cosmic shocks \citep{karyjo96}, although their confinement within cosmic
structures becomes difficult beyond $10^{7}$ GeV \citep{voahbr96,cobl98}.
The DSA process takes place on time-scales much shorter than the 
simulation time-step and, from a numerical point of view, it is 
treated as ``instantaneous''. Thus, the injected particles are
redistributed to a supra-thermal population in accord with the shock 
properties.
In the test-particle limit adopted here, this is a power law in 
momentum given by \citep[\eg][]{drury83}
\begin{equation} \label{finj.eq}
f(p)_{\rm shock}= 
f(p_{inj})_{\rm Maxwell} \left(\frac{p}{p_{inj}}\right)^{-q} .
\end{equation}
extending from $p_{inj}$ to $p_{max}$. The log-slope of the distribution is
$q\equiv \partial \ln f / \partial \ln p=3(\gamma_{gas} +1)/[2(1 - M^{-2})]=4/(1-M^{-2})$ for $\gamma_{gas}=5/3$ where $M$ is the shock Mach number and $\gamma_{gas}$ 
is the gas adiabatic index.  
The Maxwellian and the power-law components that make up the full 
distribution function join smoothly roughly at the injection 
momentum, $p_{inj}$; that is the momentum
threshold above which ions are energetic enough to escape upstream 
the shock as mentioned above. The choice of $p_{inj}$ sets the normalization 
in the expression (\ref{finj.eq}). Here, as in previous calculations, 
the momentum threshold 
is assumed as large as a few times the peak thermal value 
\citep[\eg][]{kajo95}, namely
\begin{equation} \label{pinj.eq}
p_{inj} = c_1\,2\sqrt{m_p k_{\rm B} T_{\rm shock}} 
\end{equation}
where $m_p$ is the proton mass, $k_{\rm B}$ is the Boltzmann constant.
In this study we adopt a parameter value $c_1\geq 2.5$, 
on the relatively low-efficiency 
side of canonical values. The sign, $>$, indicates that when for very strong 
shocks ($M\geq 10$) the produced CR pressure is larger than allowed by the 
test particle approximation, we additionally impose an upper limit on
it by demanding that it is $\leq 40$\% of the shock ram pressure.
This choice is admittedly artificial, although it does resemble the
effect of a CR back-reaction on the shock structure which suppresses
the injection at the sub-shock \citep[\eg][]{beel99}.
More importantly, however, it allows us to control how the 
adopted parameter affects the final result.

In order to follow the further evolution of the injected CRs we divide 
momentum space into $N_p$ logarithmically 
equidistant intervals ({\it momentum bins}) and for each point mesh 
of the spatial grid, ${\bf x_j}$, we define the following piecewise power-law
distribution function \citep{jre99,min01}
\begin{equation} \label{distf.eq}
f({\bf x}_i,p) = f_j({\bf x}_i) \, p^{-q_j({\bf x}_i)},
~~~~~~~~ 1<p_{j-1} \le p \le p_j,
\end{equation}
where $p_j ... $ are the momentum bins' extrema. 
Spatial propagation and energy losses/gains of the accelerated CRs are 
then followed by solving numerically a ``kinetic'' equation written in 
comoving coordinates and integrated over the aforementioned momentum bins,
namely \citep{min01}
\begin{equation}  \label{numden.eq}
\frac{\partial n({\bf x_j},p_i)}{\partial t} =
- {\bf \nabla\cdot } \left[ {\bf u}\,n({\bf x_j},p_i)\right] + 
\left[ 
b(p) \;4\pi \;p^2\; f({\bf x_j},p)\right] _{p_{i-1}}^{p_i}  
+Q({\bf x_j},p_i). \label{dce3.eq}
\end{equation}
Here $n({\bf x},p_i) = 
\int_{p_{i-1}}^{p_i} 4\pi\,p^2 \;f({\bf x_j},p)\; dp$ is 
the number density of CR in the $i_{th}$ momentum bin, 
$b(p) \equiv -\left(\frac{dp}{d\tau}\right)_{tot}$ describes adiabatic 
energy losses/gains as well as Coulomb collisions and p-p inelastic 
scattering and $Q_i$ is a source term, $i({\bf x_j},p_i)$, integrated 
over that momentum bin.
At each time-step, $q_j({\bf x}_i)$ is determined self-consistently 
from $n({\bf x}_i,p_j)$ and the required continuity of $f({\bf
x}_i,p_j)$ at bins interfaces \citep{jre99}.
Here we employed 4 momentum bins, although our tests indicate that 
the results are unchanged when $N_p$ is doubled.

\subsubsection{Semi-implicit COSMOCR} \label{cosrayele.se}

Injection of thermal electrons in the acceleration mechanism is here simply
modeled by assuming that the ratio between CR electrons and ions at
relativistic energies is $R_{e/p}=10^{-2}$. 
Observationally, for Galactic CRs this ratio has been 
measured in the range 
$1\times 10^{-2}-5 \times 10^{-2}$ \citep{muta87,mulletal95,alpego01}.
The introduction of this parameter \cite[see also][]{elbeba00}
simplifies the treatment of this process, which is otherwise quite complex.
In fact, the shock dynamics is presumably dominated by the proton component
with which most of the mass, momentum and energy are associated.
However, for an effective electron-wave interaction
the resonant condition requires $\omega - k_\parallel v_\parallel =
\pm \omega_{ci}$, where $\omega$ and $k$ are the wave frequency and
wave-vector respectively, $v$ is the particle speed, the symbol
$\parallel$ indicates the component parallel to the ordered magnetic 
field, $B_0$, and $\omega_{Le,i}= eB_0/m_{e,i}c$ is the
ionic/electronic ($i,e$)
Larmor frequency. Thus, wave-modes generated by leaking protons, 
with $k\sim \omega_{Li}/v$, are much lower than those with which a thermal 
electron can interact. 
Therefore, either electrons are pre-accelerated \citep[\eg][]{mcclem97} 
or a different injection model is required 
\citep[\eg][]{levinson92,levinson94,byuv99}. 

The dynamics of CR electrons is further complicated by the severe energy
losses they suffer. For this reason the numerical treatment 
is far more delicate than for the CR ions
as the distribution function steepens and cuts off 
above a certain momentum. In order to determine the 
evolution of this component more accurately,
in addition to the electron bin number density described by eq. 
(\ref{numden.eq}), 
we also follow the associated kinetic energy 
$ \varepsilon_i  =4\pi \int_{p_{i-1}}^{p_i}\,p^2 \;f(p)\,T(p)\; dp $. 
Here $T(p)=(\gamma -1)\,m_e\,c^2$ is the particle kinetic energy and
$\gamma = 1/\sqrt{1-(v/c)^2}$ is the Lorentz factor. 
The equation describing the evolution of $\varepsilon({\bf x_j},p_i)$ 
is obtained in analogy to eq. (\ref{numden.eq}), by integrating over the $i_{th}$ 
momentum bin a kinetic equation that has been multiplied by $T(p)$.
This reads \citep{min01}
\begin{equation} \label{enden.eq}
\frac{\partial \varepsilon ({\bf x_j},p_i)}{\partial t} =
- {\bf \nabla\cdot } \left[ {\bf u}\,\varepsilon({\bf x_j},p_i)\right] + 
\left[ b(p) \;4\pi \;p^2\; f({\bf x_j},p)\,T(p) \right] _{p_{i-1}}^{p_i} -
\int_{p_{i-1}}^{p_i} b(p)
\;\frac{4\pi c p^3 f({\bf x_j},p)}{\sqrt{m_ec^2+p^2}}\;dp
+ S({\bf x_j},p_i),
\end{equation}
where 
$S({\bf x_j},p_i) = 4
\pi\;\int_{p_{i-1}}^{p_i}\,p^2 T(p) i({\bf x_j},p_i)$
and the third term on the right hand side, so written for numerical reasons,
includes a combination of contributions from CR 
pressure work and sink terms. Thus the evolution of the CR electrons is 
described by eq. (\ref{numden.eq}) and (\ref{enden.eq}) and now the slope 
of the distribution function at each grid point and momentum bin is
determined self consistently by the values of 
$n({\bf x_j},p_i)$ and $\varepsilon({\bf x_j},p_i)$.
In addition, at each grid zone we also follow explicitly the 
upper momentum cut-off in the electron distribution function.

For an accurate time integration of the system of eq. 
(\ref{numden.eq}) and (\ref{enden.eq})
the {\it explicit} formulation described in \cite{min01} required 
a time-step, $\Delta t$, much shorter than the electron 
cooling time, $\tau_{cool}$.
And in fact, in order to properly compute in a cosmological simulation 
(where $\Delta t \sim 10^7-10^8$yr) the electrons responsible 
for synchrotron and inverse 
Compton emission at the frequencies of interest,
\cite{mjkr01} had to appropriately subcycle the integration of 
eq. (\ref{numden.eq}) and (\ref{enden.eq}) for several iterations.
This procedure is not computationally convenient 
particularly at high red-shift.
Furthermore, it becomes impractical
when, in order to compute cosmic background radiation as in the
present instance, the CR electron population is needed throughout
most of the simulation.
For this reason, for the electron population
we have reformulated the momentum space portion 
(\ie without the advection term)
of the numerical scheme in a semi-implicit form. 
Such reformulation is 
quite straightforward an exercise and it mainly consists of expressing 
all the contribution to the fluxes in momentum space in terms 
of $n({\bf x_j},p_i)$ and $\varepsilon({\bf x_j},p_i)$ 
at times $t$ and $t+\Delta t$. The semi-implicit formulation is 
nominally less precise than the explicit one, 
although this will only affect the 
region near the momentum cut-off. Here, however,
even though we explicitly follow the upper momentum cut-off,
the major limitations of the numerical model are due to the sub-grid 
power-law description of the distribution function
rather than to the semi-implicit formulation.
Therefore, there is no significant loss of accuracy with our new approach. 

Given the finite simulation time-step, the time evolution of 
the very high and very low energy ends of the distribution function,
characterized by a cooling time $\tau_{cool}(p)<\Delta t$, 
cannot be followed even with a semi-implicit scheme.
We notice that in the transport equation the diffusion term can be
neglected away from shocks \citep{min01,jre99}
and CRs are basically transported by the fluid (to which 
they are coupled through the magnetic field) at the flow speed $v$. 
Thus, because of the Courant condition, $v\Delta t \leq \Delta x$, 
where $\Delta x$ is the mesh size, electrons such that
$\tau_{cool} (p)< \Delta t$, will not propagate outside 
the grid zone in which they have been created. 
Therefore, in these energy ranges 
it is appropriate to take the steady state solution to eq. 
(\ref{numden.eq}) as
\begin{equation}
n({\bf x_j},p_i) = 4\pi\int_{p_{i-1}}^{p_i}\,p^2 \; 
f_{c}({\bf x_j},p_i)\; dp  =
- 4\pi \int_{p_{i-1}}^{p_i} 
\frac{dp}{b(p)}\; \int_{p}^\infty \; \rho^2 \,i({\bf x_j},\rho)\, d\rho .
  \label{sss.eq}
\end{equation}
\noindent
We caution that for the case of shock acceleration, 
the above steady-state solution expresses
the balance between a source and a loss term within the volume of 
a grid zone, but not at the acceleration region. 
In fact, the rate of energy gain at a shock must be higher than the loss rate 
up to the maximum energy of the accelerated particles. 
Physically, the expression in eq. (\ref{sss.eq}) represents a 
{\it cumulative population} $f_c$, \ie
a summation of all the individual populations of CR electrons within 
the grid zone ${\bf x_j}$ that, after they emerge from the acceleration 
region, and as they are being advected away from the shock, start {\it aging}.
Aging here refers to the modifications produced by energy losses.
It is easy to infer from eq. (\ref{sss.eq}) that if the source 
term is a power law 
with slope $q_s$, the distribution $f_c$ will have an energy dependent slope
\begin{equation} \label{slop.eq}
q_c(p) = \frac{\partial \ln f_c}{\partial \ln p} = 
q_s - 1 + \frac{\partial \ln b(p)}{\partial \ln p}.
\end{equation}
\noindent
For dominant synchrotron and inverse Compton losses, as in the case of very 
high energy electrons, eq. (\ref{slop.eq}) gives $q_c\simeq q_s+1$,
whereas when Coulomb collisions dominate $q_c\simeq q_s-1$,.
Thus, effectively because of cooling and of a finite 
computational time-step, the 
numerical solution at a shock will be a broken power-law. At high energy
the break will occur at momentum
\begin{equation}
p_{break}^h \sim 10^4 \left(\frac{\Delta t}{10^8 {\rm yr}}\right)^{-1}\; 
\left(1+\frac{\rm U_B}{\rm U_{CMB}}\right)^{-1} {\rm MeV},
\end{equation}
\noindent
where ${\rm U_B}$ and ${\rm U_{CMB}}$ refer to the magnetic and 
CMB energy density respectively. 
At low energy, the break corresponds to
\begin{equation}
p_{break}^l \sim 1.8 \left(\frac{\Delta t}{10^8 {\rm yr}}\right)\; 
\left(\frac{n_{\rm gas}}{10^{-3}{\rm cm}^{-3}}\right) {\rm MeV}.
\end{equation}
\noindent
For the primary CR electrons we use 5 logarithmically equidistant
bins, spanning momentum space from $p_{min}=15$ MeV up to 
$p_{2}= 10^2$ GeV. As for the case of CR ions, our tests show that 
the final results are not affected in any significant way by increasing
the number of momentum bins. 
In addition, the distribution function above $p_2$, computed according 
to the ``steady state'' solution [eq. (\ref{sss.eq})] for zones with
non-null source terms, 
is recorded in an extra momentum bin ranging from $p_2$ up to 
$2\times 10$ TeV. The latter is an appropriate value for the maximum energy of
accelerated CR electrons, provided a magnetic field 
of order $0.1 \mu$G throughout the IGM \citep{lowa00}.

Although not needed in the present study, we also have the capability
of computing the sub-relativistic component below $p_{break}^l $.
This component might be of interest, for example in order 
to estimate the emission of high energy radiation 
through non-thermal bremsstrahlung \cite[\eg][]{min02a}.
Notice that in our model the energy associated to both CR ions 
and electrons is always fraction of the shock ram pressure.
The small fraction of energy that is carried
by the sub-relativistic components of both these CR populations,
with time is transferred back to the gas in thermal form through
Coulomb collisions. However, this same energy would
have been converted into thermal energy through dissipation at
the shock anyway, had DSA not be present. Therefore, it is obvious that
the sub-relativistic populations created in our model do not 
produce any ``thermodynamic catastrophe''
\citep[\cf][ and references therein]{petro01}.

\subsubsection{Secondary Electrons and Positrons}

At each time-step we also compute the distribution of high energy
\gr emitting secondary electrons and positrons (e$^{\pm}$) 
produced in hadronic collisions
of CR ions with the nuclei of the intra-cluster gas, within the 
same momentum range as for the shock accelerated electrons.
Secondary $e^\pm$ are generated in the decay 
of charged muons according to
\begin{eqnarray} 
\mu^\pm  \rightarrow e^\pm + \nu_e (\bar{\nu}_e) + \bar{\nu}_{\mu}(\nu_{\mu})
\end{eqnarray}
\noindent
which in turn are produced in the following reactions
\begin{eqnarray} 
p + p & \rightarrow \pi^\pm +  X,~~ & \pi^\pm  \rightarrow \mu^\pm + \nu (\bar{\nu}) \\
p + p & \rightarrow K^\pm + X, ~~ & K^\pm  \rightarrow \mu^\pm +\nu (\bar{\nu})\\
& & K^\pm  \rightarrow \pi^0 + \pi^\pm
\end{eqnarray}
\noindent
In addition to $p+p$ inelastic collisions, 
the above cascades are also triggered by the following interactions between CR
ions and thermal nuclei: $p+$He, $\alpha +$H and $\alpha +$He. We include them 
by assuming a helium number fraction
of 7.3\% for the background gas and a ratio
$(H/He) \simeq 15$ at fixed energy-per-nucleon 
for the CRs \citep{medrel97}. Further details on the 
source term for secondary $e^\pm$ are given in \cite{min01}.


\section{Results} \label{res.se}

The shock strength, which we classify according to the 
shock Mach number, $M$, affects the CR produced background radiation
in two respects: the normalization, which turns out higher for 
a flatter distribution function because it allows for a larger number
of high energy, 
\gr emitting CRs; and the spectral index which, as shown below,
has a simple dependence on the distribution function log-slope $q$.
In addition, we notice that for a given injection rate, high Mach number 
shocks are more efficient in converting the shock ram pressure 
into CR ion pressure.
As a first step, therefore, we investigate the strength  of the 
shocks that were chiefly responsible for the heating of the \igm and from which 
the primary CRs originate.

The Mach number is given by the ratio of the shock speed to the 
local sound speed \citep{lali6}.
When computing the shock Mach number we assume that the 
pre-shock gas temperature is not less than $ 10^4$ K, 
the value it would quickly reach had photo-heating due to 
UV background light as well as radiation from  
the hot post-shock region been included. For this reason,
the highest shock Mach numbers reported below do not 
exceed a few 100, contrary to the case in \citet{minetal00} where 
such lower limit on the gas temperature had not been assumed.

\subsection{Inter-galactic Shocks} \label{igms.se}

\begin{figure}
\includegraphics[width=0.50\textwidth ]{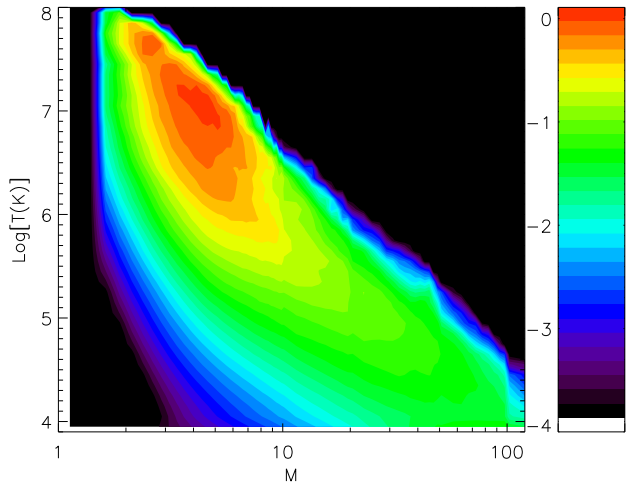}
\caption{\textit{Top: Two dimensional diagram showing the differential
$\frac{\partial^2\Delta E_{th}}{\partial\log M\partial \log T_1}$ as a
function of pre-shock temperature, $T_1$, and shock Mach number, $M$.
$\Delta E_{th}$ is defined in eq. (\ref{thermen.eq}). It represents
integrated thermal energy produced
at cosmic shocks during cosmic history. It is shown in units of
keV per particle.}}
\label{shfl2.fig}
\end{figure}
The amount of thermal energy produced at
shock fronts throughout cosmic history is given by the expression
\begin{equation} \label{thermen.eq}
\Delta E_{th}  =
\int_{\tau _i}^{\tau (z=0)} dt \int dS_{shock} k_B \Delta T (n v)_{shock}
\end{equation}
where $t$ is cosmic time, $dS_{shock}$ is an element of shock surface 
and $(nv)_{shock}$ and $\Delta T$ the gas flux and temperature 
jump across it. Thus, the quantity $dS_{shock} k_B \Delta T (n
v)_{shock}$ represents the amount of kinetic energy dissipated at the 
shock into thermal energy per particle and per unit time. 
In order to determine which shocks were primarily responsible for 
heating the IGM, in Fig. \ref{shfl2.fig} we plot the
differential quantity 
$\frac{\partial^2\Delta E_{th}}{\partial\log M\partial \log T_1}$ as a
function of pre-shock temperature, $T_1$, and shock Mach number, $M$.
It shows that in terms of parameter space $(M,T)$, shock heating is
confined to a narrow strip whose upper and lower edges are loci
of constant velocity of order a few $10^3 \kms$ and $10^2 \kms$ 
respectively.
A region where $3<M<10$ and $10^6 < T < {\rm a ~few} \times 10^7$ further
stands out where shock heating seems most ``influential''.
More specifically,  from a quantitative analysis we find that thermal 
energy
due to shock heating is contributed in the amount of 30\% by weak shocks
($M<4$), 45\% by moderately strong shocks ($4\leq M \leq 10$)
and 25\% by strong shocks ($M>10$).
Thus we expect most of the CRs accelerated at \ig shocks to be described
by flat distribution functions.

\subsection{The \gr Background}

The \gr flux in units of `keV cm$^{-2}$ s$^{-1}$ sr$^{-1}$' 
at a given photon energy, $\varepsilon$, is computed as
\begin{equation}
\varepsilon^2 J(\varepsilon)
= \varepsilon \, \frac{c}{4\pi H_0} \; \int_{0}^{z_{max}}
\frac{e^{-\tau_{\gamma\gamma}}}{[\Omega_m (1+z)^3 + \Omega_\Lambda]^{1/2}} \;
\frac{j[\varepsilon (1+z),z]}{(1+z)^4} \; dz
\end{equation}
where $j(\varepsilon ,z)$ is the computational-box-averaged
spectral emissivity in units `photons cm$^{-3}$ s$^{-1}$'
computed at each simulation
red-shift, $z$, and at the appropriately blue-shifted photon 
energy $\varepsilon (1+z)$. In addition, $\tau_{\gamma\gamma}$
is an attenuation factor due to photo-pair creation,
$\gamma\gamma \rightarrow e^\pm$. It can actually be 
ignored given that the energy, $\varepsilon$, of the measured \gr 
photons is below 100 GeV and that the contribution from the
high red-shift universe is not significant (see below). 
Finally, $c$ indicates the speed of light and
$z_{max}$ an upper limit of integration.
The actual value of the latter parameter 
is not too important as most of the \gr emission is produced 
in the low red-shift universe, with $j(\varepsilon,z)$ decreasing 
by about an order of magnitude per unit red-shift increase. 
The strong red-shift evolution of $j(\varepsilon ,z)$ is related to 
the available amount of kinetic energy to be dissipated at shocks,
which is itself a strongly decreasing function of red-shift 
\citep{minetal00}.
We consider two emission processes: \ic of CR electrons scattering off
CMB photons (leptonic), and decay of neutral pions, 
$\pi^0 \rightarrow \gamma\gamma$ (hadronic),
produced in p-p inelastic collisions of CR ions off nuclei 
of \ig gas. In addition we also consider \ic emission 
from secondary \epm produced in the aforementioned p-p collisions.
Our treatment of \ic emission and $\pi^0$ production are detailed
in \cite{mjkr01} and  \cite{mrkj01} respectively. For \ic
emission we simply use a Thompson cross section, as
Klein-Nishina corrections become relevant only for 
electron energies about an order of magnitude above the maximum
considered here.
\begin{figure}
\includegraphics[width=0.50\textwidth ]{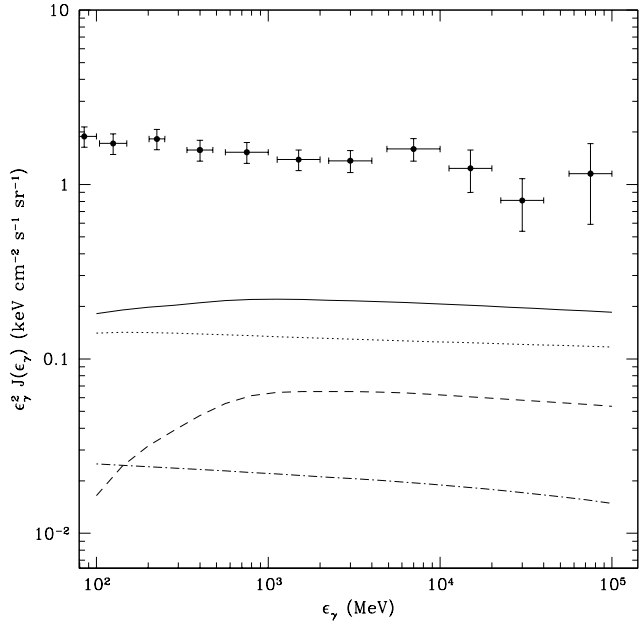}
\caption{\textit{Top: The total
\gr background flux produced by 
cosmological CRs (solid line).
It includes the contributions from \ic emission from
shock accelerated CR electrons (dotted line),
decay of $\pi^0$ generated in inelastic p-p collisions
(dashed line) and \ic emission from secondary \epm 
(dashed-dotted line). The experimental data and their error-bars 
(solid dots) are taken from \citet{sreeku98}.}  }
\label{gamma.fig}
\end{figure}
The results are shown in Fig. \ref{gamma.fig} where we plot 
$\varepsilon ^2 J(\varepsilon)$ 
as a function of $\varepsilon$. The total
flux (solid line) corresponds to a constant value at the level of 
0.2 keV cm$^{-2}$ s$^{-1}$ sr$^{-1}$ throughout the spectrum. 
It is clearly dominated by \ic emission from
shock accelerated CR electrons (dotted line). A fraction of 
order 30\% is also produced by $\pi^0$-decay (dashed line), 
whereas \ic emission from secondary \epm (dotted-dashed line) 
turns out about an order of magnitude smaller, although it is of the same order
as $\pi^0$-decay and becomes larger than it around 100 MeV.
For comparison the observational data, taken from the EGRET experiment
\citep{sreeku98}, are also plotted with the associated error-bars (solid dots).

Remarkably, all three components produce the 
same flat spectrum, similar in shape to the observed one,
except below $\sim $ GeV where the $\pi^0$ production 
cross section drops due to phase-space constraints.
This result is a reflection of the fact that the CRs
at the origin of the \gr radiation were produced in strong
shocks and have a flat distribution function
with a log-slope $q\simeq 4$. Indeed, 
\ic emissivity from a power-law distribution of CR electrons, 
$f(p)\propto p^{-q_e}$, is of the form $j(\varepsilon)\propto
\varepsilon^{-(q_e-3)/2}$.
Therefore, for a steady state distribution
function of CR electrons accelerated at strong shocks ($M>4$)
and subject to \ic losses $q_e\simeq 5$ [see eq. (\ref{slop.eq})], 
$\varepsilon ^2J(\varepsilon)
\propto \varepsilon j(\varepsilon) \propto \varepsilon^0$. 
As for the hadronic component, 
the pion emissivity (in units of pions s$^{-1}$ cm$^{-3}$) 
from a CR ion distribution $f(p)\propto p^{-q_p}$, 
is $j_{\pi^0}(\varepsilon_{\pi^0}) 
\propto \varepsilon_{\pi^0}^{-(4q_p-13)/3}$ \citep[\cf][]{masc94}.
Thus, for $q_p\simeq 4 $ (strong shocks), as before
$\varepsilon ^2 J(\varepsilon)
\propto \varepsilon^0 $. Analogously, the secondary \epm injected 
in the above hadronic processes reach a steady state distribution 
against \ic losses of the form $f_{e^\pm}(\varepsilon_{e^\pm})
\propto \varepsilon_{e^\pm} ^{-(4q_p-1)/3}\propto
\varepsilon_{e^\pm}^5$ for $q_p\simeq 4$ 
[\cf eq. (\ref{slop.eq}) and \citealt{masc94}].
As the steady state distribution of primary $e^-$, this component
also produces an \ic emission $j(\varepsilon)\propto \varepsilon^{-1}$
and a \gr background flux $
\varepsilon j(\varepsilon) \propto \varepsilon^0$.

Although with a spectral shape that accords with observations, 
the computed \gr flux is only 15 \% of the observed CGB. 
A number of factors, which we address below, might affect these results.
Firstly, resolution effects could be important,
particularly for the hadronic component, because neutral 
$\pi$-mesons 
are produced in a two body process. In fact the emission
from smaller structures is likely to be underestimated, although 
we do not expect the hadronic component to be dominated by 
the contribution of the low end of the mass function. This is corroborated
by the fact that our present estimate for the flux produced by this process
is already larger than, although of the same order of magnitude as, 
the upper limit given by the analytical calculations carried out by 
\cite{cobl98}.
In particular, between 1 GeV and 100 GeV the computed \gr flux from \pnd 
corresponds to about 5 \% of the measured value and to $\sim$ 7\% when
combined with the contribution of \ic emission from $e^\pm$.
In comparison \citet{cobl98} estimated a range of values $0.5- 2$ \% 
(but they did not include \ic emission from secondary $e^\pm$).
Thus the error on the \gr flux as a result of hadronic processes 
should amount to a factor two or so. 
As for shock accelerated electrons, our tests indicate that coarse 
grid effects are not important and that they amount at most to a few tens of 
percent.

More important is the issue related to the assumed acceleration efficiency,
$\theta$, for converting the shock ram pressure into CR pressure.
In this calculations we adopted $\theta_{max} \leq 40\%$ with only a
small fraction $R_{e/p}\simeq 10^{-2}$ of the generated CR pressure to be
borne by electrons.
Shock acceleration can be even more efficient than allowed here.
However, in those cases we expect non-linear effects to suppress the 
injection rate with consequent reduction of the \gr emitting CR population
whilst most of the CR pressure is carried by the upper end of a concave-up
distribution of CR ions. 
In principle the parameter $R_{e/p}$ could be higher, allowing for a 
substantially larger contribution from the leptonic component.
This point is particularly important because it has been recently claimed 
that IC \gr flux from CR accelerated at \ig shocks can make about all of 
the unexplained CGB \citep{lowa00}. 
Therefore, in the following we will try to assess
the efficiency of conversion of shock ram pressure into CR electrons
based on EGRET observations of nearby clusters. 

\subsection{IC \gr Emission from Clusters of Galaxies}

For this purpose,
we have computed the integrated photon 
\gr luminosity above 100 MeV, L$_\gamma(>100$ MeV), 
produced by primary CR electrons accelerated 
at accretion shocks associated to groups/clusters of galaxies. 
L$_\gamma$ was computed within a volume of radius 
about 5 \hinv Mpc around the cluster center, although most of the flux is
produced roughly within a radius of 3 \hinv Mpc \citep{min02a}. 
Collapsed objects were identified through 
a slightly modified version of the ``spherical over-density'' method
\citep{laco94}, fully detailed in \citet{minetal00}. 
However, because of the large spatial extension of the \ic emission,
there is a real risk of mis-association with any selected object of 
radiation emission produced in nearby but unrelated accretion shocks.
To circumvent this problem we further refined our sample by
rejecting the smaller of two or more objects that happen to 
be closer to each other than a distance of 7.5 Mpc.
\begin{figure}
\includegraphics[width=0.50\textwidth 10]{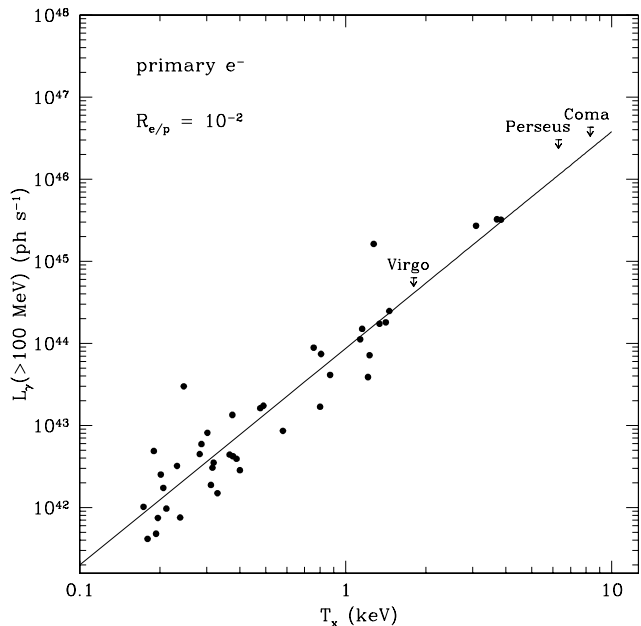}
\caption{\textit{Top: The \ic \gr photon luminosity above 100 MeV
from individual clusters as a function of the cluster 
X-ray luminosity weighted core temperature. The flux
is produced by CR electrons accelerated 
at accretion shocks associated to groups/clusters of galaxies.}  }
\label{grtx.fig}
\end{figure}
In fig. \ref{grtx.fig} we plot 
the computed \gr photon luminosity, L$_\gamma (>100~ {\rm MeV})$,
against T$_x$, that is the thermal X-ray luminosity weighted cluster 
core temperature.
The correlation between the plotted quantities is 
tight and the numerical data are well fit with a single power-law
function. After performing a simple least-$\chi ^2$ analysis we 
obtain the best-fit parameters as
\begin{equation} \label{gamfit.eq}
{\rm L}_\gamma (>100~ {\rm MeV})= 8.7\times 10^{43}  
\left(\frac{R_{e/p}}{10^{-2}}\right)
\; \left(\frac{\rm T}{\rm keV}\right)^{2.6} \; {\rm ph ~s}^{-1}.
\end{equation}
The temperature scaling for L$_\gamma$ obtained is close to
what expected based on clusters scaling relations. 
In fact, in a steady state configuration appropriate here due 
to the short CR electrons cooling time, the radiation flux
must be produced at the same rate at which kinetic energy is 
being supplied to the external accretion shocks. That implies 
$F_\gamma \propto \rho v_{shock}^3 R^2 \sim T_x^{2.5}$. 
The plot in fig. \ref{grtx.fig} is noteworthy by itself.
In fact it shows that the \gr flux associated to \ic emission from
shock accelerated electrons is comparable with the flux from 
$\pi^0$-decay which is usually regarded as the main 
source of $\gamma$-rays from clusters.
Thus, for a correct interpretation of the \gr emission from
clusters that planned \gr facilities aim at detecting, 
this component needs to be properly addressed.
This task is being undertaken in a separate paper \citep{min02a}.

In order to constrain the electron acceleration efficiency 
in fig. \ref{grtx.fig} we compare the computed \gr photon luminosity
above 100 MeV to upper limits set by the EGRET experiment
on the \gr flux from nearby clusters of galaxies. 
Of the 58 cases that appear in the list by \cite{reimer99},
we found that the tightest constraints were those set by the
experimental upper limits on the following
clusters: Virgo, Perseus and Coma. The latter two have temperatures
larger than those of objects found in the computational box and,
therefore, their upper limits must be compared to values extrapolated
from the the scaling relation given in eq. (\ref{gamfit.eq}).
The flux upper limits were converted into luminosity limits 
based on the relation L$_\gamma = 4\pi d_L^2$F$_\gamma$, where $d_L$ is the
luminosity distance taken, for each cluster, 
from the literature. In table \ref{tab1.ta}
we summarize the assumed values for the cluster temperature, 
distance and \gr flux upper limits. The observational data adopted
here, although numerically very close to those published in 
\cite{reimer99}, were recently re-obtained by Reimer after 
careful and improved analysis of numerous staked observations 
for each source (Reimer, priv. comm.). 
\begin{table}
\caption{}
\label{tab1.ta}
\begin{tabular}{lcccc}\hline \hline
 Object & T$_x$ & d$_L$ & F$_\gamma(>$100 MeV) $^{\rm a}$         & References  \\
        & keV   & Mpc   &  ph cm$^{-2}$ s$^{-1}$ &             \\
\hline
Virgo \hfill    & 1.8 & 15.5 & $<2.20 \times 10^{-8} $ & 1,2,3 \\
Perseus \hfill  & 6.3 & 82.5 & $<3.73  \times 10^{-8} $ & 4  \\
Coma \hfill     & 8.3 & 104.9 & $<3.26  \times 10^{-8} $ & 5,6  \\
\hline
\end{tabular} 
\smallskip
\begin{list}{}{}
\item[$^{\rm a}$] {From O. Reimer, priv. comm.}
\item[References -] (1) \citealt{grahametal99}; (2) \citealt{fssb01};
(3) \citealt{bohrin94}; (4) \citealt{schwarzetal92}; 
(5) \citealt{arnaudetal01}; (6) \citealt{baumetal97};\\
\end{list}

\end{table}
The plotted experimental upper limits (particularly those
relative to Coma and Virgo clusters) require that the 
fraction of shock energy that goes into relativistic CR electrons 
be $\leq $0.8\% $\times [\log(p_{max}/m_ec)/\log(4\times 10^7)]$
where $p_{max}$ is the maximum energy of the accelerated electrons. 
For the adopted ion injection efficiency this roughly translates into
$R_{e/p} \leq 2 \times 10^{-2}$.
In any case, this allows us to set an upper limit on the computed 
\gr flux of about 0.35 keV cm$^{-2}$ s$^{-1}$ sr$^{-1}$.
Thus, according to our computation, cosmological CRs could overall
contribute an important fraction of order $\sim 25$ \% of the 
extra-galactic 
cosmic \gr background as measured by EGRET \citep{sreeku98}. 
Due to the scatter in the plot of fig. \ref{grtx.fig} - which we regard 
as ``real'' and attribute to the different shock histories of the simulated 
objects -  it is plausible to relax the upper limit on the 
acceleration efficiency of CR electrons by another factor $<2$.
This is allowed, however, only if the \gr
flux produced by \pnd  is negligible.
And although this is an extreme scenario, it cannot be 
excluded because $\gamma$-rays from galaxy clusters/groups 
have never been measured. Thus in this latter case the 
\ic \gr flux could account for up to $\sim 30$ \% of the measured CGB.
Notice that the upper limits derived here rely mainly on the assumption 
that we compute correctly the relative fraction of \gr flux from accretion 
shocks onto clusters and on smaller structures, where most of the emission 
actually originates. 

To summarize, the whole unexplained \gr background flux could be 
accounted for by assuming an acceleration efficiency for the electronic 
component $\sim 3 \%$, or higher by a factor $\sim 7.5$ than assumed
in fig. \ref{gamma.fig}, but that would result in fluxes from nearby clusters 
such as Virgo, Coma and Perseus well above the experimental upper limits.
It is instructive to compare in some more detail our approach  
and that of \citet{lowa00} as the differences go beyond the 
assumed acceleration efficiency for the electronic component. 
In fact, in \lowa model, the population of 
emitting CR electrons is computed by postulating that a 
fraction $\xi_e$ of the thermal energy in the shocked IGM is 
converted into a flat power-law of relativistic electrons. 
The thermal energy density of the shocked IGM is estimated 
as $1.5 f_{sh} n_p k_B T$, where $f_{sh}$ is the fraction of 
shocked baryons, $n_p$ the average proton number density and $T$ 
the mass-weighted average gas temperature. For their quantitative 
estimates \cite{lowa00} assume $f_{sh} (k_BT/1{\rm keV})\sim 1$ 
and $\xi_e \simeq 0.05$. 
Notice that, unlike $\xi_e$, the efficiency parameter defined in 
our case is based on the shock ram pressure instead of the 
post-shock thermal energy density.
For moderately strong and strong shocks, 
the two definitions differs by 12.5 \% (it gets more complicated 
for weak shocks but this case is not relevant because weak shocks 
are inefficient accelerators). In particular, assuming $\xi_e \simeq 0.05$ 
as in \cite{lowa00} means that 5.6 \% of the ram pressure is converted
into CR electrons. This corresponds to a factor $\sim 15-19$  larger than
the value of 0.3-0.4 \% adopted to compute the dotted line in fig. 
\ref{gamma.fig} which lies, however, only a factor 10 below \lowa's 
estimate. This difference by a factor $\sim 1.5-2$
grows if one accounts for the following additional matters:
effectively in the simulation $f_{sh} \sim 2/3$ and $k_B T/keV \sim 1/2$ 
\citep[see also][]{keshetetal02}. In addition, as pointed out 
in \S \ref{igms.se},
only $\sim 2/3$ of the IGM thermal energy comes from shocks strong 
enough (\ie $M\geq 4$) to produce \gr emitting CR electrons. 
Thus, using the appropriate fudge factors \lowa expression would 
produce a result smaller than the simulation by almost an order
of magnitude (factor $\sim 7$). 
In part this is due to the 
fact that, by following the full evolution of the shock accelerated 
CR electrons from 10 MeV to 2 TeV, 
we are also able to account for a re-acceleration process.
That is the re-energization of {\it aged} (by cooling) CR electrons
that were produced in earlier shock events.
As already pointed out, this effect increases our computed CGB 
by a factor between 2 and 3. Part of the remaining discrepancy 
could be due to \gr flux emitted at high red-shift ($z>0.5$) 
that \lowa neglected,
although according to our simulation that should not be too significant.
Most likely, however, it reflects the fact that the actual conversion 
of shock energy into CR electrons cannot be simply estimated through the 
IGM average temperature. In any case, given its simple formulation 
that estimate is quite close to our results.

Just prior to submission of this paper, the work by \cite{keshetetal02} 
was brought to our attention. 
The authors further investigate the \cite{lowa00} model utilizing
smoothed particle hydrodynamic (SPH) numerical simulations. 
They detect the occurrence of shocks in individual SPH particles 
by measuring changes in their entropy every some time interval.
When shocks are identified a fraction $\xi_e$ of the generated 
post-shock thermal energy is converted into CR electron energy.
The CR electrons are distributed in energy according to the test particle
limit of DSA theory, as assumed here.
Their results, particularly the level of CGB produced by shock accelerated
electrons, is in good agreement with our estimated flux plotted 
in fig. \ref{gamma.fig}.
However, there are some differences that are worth pointing out.
In fact, for the same estimated \gr flux our simulation requires an 
electron acceleration efficiency smaller than theirs by about 
an order of magnitude (of course this includes already the 
different definitions of efficiency).
We find that most of the thermal energy in the low red-shift
universe was produced by what we refer to as moderately strong 
shocks, that is shocks with Mach number in the range 4-10.
Those authors mention some difficulty in tracking these shocks.
If these are neglected according to our results in \S \ref{igms.se}
one would be considering only about 1/3 of the total flux through 
shocks (although their estimate indicates more of a factor 1/2).
In addition our simulation also includes a re-acceleration process which, 
as mentioned before, boosts the computed flux by a factor 2-3.
These two factors, namely that related to moderately strong shocks and
re-acceleration, are probably the most important in determining the 
differences between our and Keshet et al. results.
\begin{figure}
\includegraphics[width=0.50\textwidth 10]{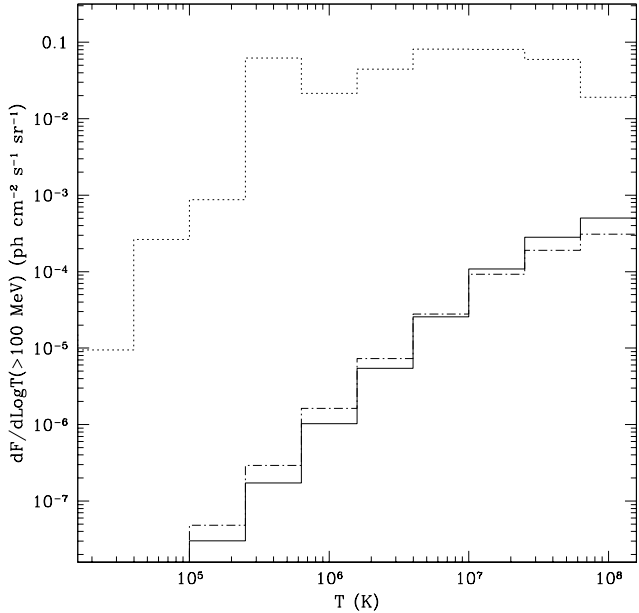}
\caption{\textit{Top: Histogram of the integrated
photon flux above 100 MeV against temperature of the
IGM where the emission originates. We show the case
for \ic emission from shock accelerated CR electrons
(dotted line), for emission from $\pi^0$-decay (solid line)
and for \ic emission from secondary \epm (dashed line)}.  }
\label{grhist.fig}
\end{figure}
\subsection{Emission Isotropy}

In order to assess the spatial distribution of the \gr emissivity, in 
fig. \ref{grhist.fig} we show a histogram of the total integrated
photon flux above 100 MeV, F$_\gamma(>100$ MeV)= $\int_{\rm 100 MeV}
J(\varepsilon) d\varepsilon $, 
versus the temperature of the \igm where it originates for the 
three computed components. The diagram illustrates that
the contributions from both $\pi^0$-decay (solid line) and
secondary \epm (dashed line) are strongly concentrated within 
regions with temperature $T\ge 10^7$keV, corresponding to groups 
and clusters of galaxies. On the contrary the 
\ic emission by shock accelerated CR electrons (dotted line)
is roughly equally
produced in the shocked IGM with temperatures ranging from 
a few $\times 10^5$ K up to several  $\times 10^7$ K. 
Notice, however, that because of strong radiative losses,
these high energy electrons are almost exclusively 
found in ``narrow'' post-shock regions. 
Thus for this component (dotted curve), the temperature axis in 
fig. \ref{grhist.fig} is indicative of the gas temperature behind
shocks, rather than anywhere throughout the diffuse IGM.
In any case, the radiation flux produced by this component 
is expected to be highly isotropic \citep{walo00} and,
as it originates from the shocked diffuse IGM, 
if detected will allow an independent lower limit 
on the cosmic baryon density \citep{lowa00}.

For all components, the \gr surface brightness will turn out 
most prominent toward galaxy clusters even for the \ic emission.
\cite{walo00} predict for the IC component fluctuations of the 
\gr background intensity of order 40 \% on the sub-degree scale.
This fluctuations should actually be even larger when the 
hadronic component that they neglected is accounted for. In fact,
the latter should be apparent only toward massive clusters
and should dominate the emission within the cluster core
\citep{min02a}.

\section{Discussion} \label{disc.se}

We have constrained the energy acceleration efficiency 
of relativistic electrons to 
$\la 1 \times [\log(p_{max}/m_ec)/\log(4\times 10^7)] ~ \%$ 
based on EGRET experimental upper limits on the photon flux
above 100 MeV from nearby clusters. 
We specify that the found efficiency limit is valid only 
insofar as the accelerated CR electrons are power-law 
distributions extending above 
$p_{max}\sim \sqrt{75 {\rm MeV} / \langle
\varepsilon_{\rm CMB}\rangle } $ m$_e$c  $\sim 100$ GeV,
where $\langle\varepsilon_{\rm CMB}\rangle$ 
is the average CMB photon energy. 
And, in fact, strictly speaking
the found constraint actually regards the total number of CR electrons 
in this energy range.
This upper limit on the electron acceleration efficiency 
plays a crucial role in the determination of the allowed 
contribution from cosmological CRs to the CGB.
One could doubt that the EGRET measurements would apply
for extended sources such as galaxy clusters accretion shocks,
which measure several degree on the sky. 
However, the EGRET field of view is much larger than this 
and the exposure times drop by only a factor of two at about 
18$\degr $ from the instrument axis. Thus, had there been any
extended flux above the upper limit set by the EGRET experiment
they would have been detected (O. Reimer priv. comm.).

The upper limit just obtained on electron acceleration efficiency
strictly speaking concerns the acceleration at cluster
shocks, whereas a significant part 
of the \gr emission is produced in inter-galactic shocks.
However, since clusters of galaxies are the deepest potential wells
in cosmic environment, their associated shocks are indeed the
strongest.
Therefore, if anything, the emissivity is expected to be 
highest there, unless these shocks are strongly modified by a
back-reaction of CR pressure. 
In addition, unlike for massive clusters, 
shocks from accretion flows on filaments and smaller structures 
in general could be weakened by energy input into the IGM from 
supernovae, stellar winds and radio-galaxies. 
These non-gravitational processes are thought to raise the temperature
of the IGM up to $10^7$K, as it seems required by the
existence of an ``entropy floor'' in the scaling relation of 
groups of clusters \citep{pocana99}. In that case, several 
inter-galactic shocks would be suppressed and
the \gr background flux generated there substantially
reduced \citep{toin02}. From our fig. \ref{grhist.fig} 
we infer a correction factor with respect to our estimate 
of order $\sim 1/3$.
However, the actual role of such feed-back processes
is still highly controversial 
\citep[\eg][and references therein]{krye00} and alternative
scenarios in which the presence of cooling reduces dramatically the
amount of required pre-heating are being investigated 
\citep[][and references therein]{vobr01}.

According to our result in fig. \ref{grtx.fig}, the
the Gamma-ray Large Array Space Telescope should be able to
measure the \gr flux due to IC emission from CR electrons accelerated
at cluster accretion shocks \cite[see also][]{min02a}. 
In a previous work we explored with the same numerical 
technique adopted here the role of cluster accretion shocks 
as sites where the CR electrons that power 
radio relics (and not radio halos) are being produced \citep{mjkr01}.
Our numerical model was able to account for several
observed emission properties of radio relics 
\cite[see also][for direct observations
related to this model]{ebkk98,robust99}.
Interestingly the upper limit on the 
electron acceleration efficiency 
found here, corresponding to a parameter
$R_{e/p}\sim 2\times 10^{-2}$, is in line with the values 
adopted in that work in order to reproduce the measured
radio emission for a number of sources. 
This number is also in agreement with 
recent observations of supernova remnants by \cite{alpego01},
who find $R_{e/p}\sim 1/160$.
In any case, upcoming observations both in soft and hard 
\gr should allow us to put better constraints on such
parameters for clusters accretion shocks.

Thus, with these constraints on the acceleration parameters
the contribution to the CGB from cosmological CRs is estimated
at the level of 20 \% and most likely below 30 \% of the measured flux.
We point out that according to recent detailed calculations
of \cite{mostre00} the diffuse continuum \gr emission from 
the Galaxy could be significantly higher than previously 
estimated, thus lowering the actual level of the diffuse 
extra-galactic CGB.
Likely, the next generation of high sensitivity \gr observatories will 
be resolve part of the CGB into a faint population of AGNs. 
In any case, the flux contributed by 
cosmological CRs will perhaps turn out the only truly 
diffuse extra-galactic \gr background component. 

\section{Conclusions} \label{con.se}

We have investigated numerically the contribution
to the CGB from cosmological CRs. We carried out a simulation
of structure formation in the canonical $\Lambda$CDM universe
and followed directly the acceleration, spatial transport and
energy losses/gains of three different CR components: 
shock accelerated ions and electrons as well as secondary \epm. 
Our results can be summarized as follows:
\begin{itemize}
\item Cosmological CRs produce a significant fraction of order of 20 \%
and no more than 30 \% of the measure CGB. 
The computed flux is dominated by a leptonic 
component ($\sim$ 70 \%) and accounts for a non negligible 
contribution from hadronic processes ($\sim$ 30 \%). 

\item A higher flux, although in principle admissible,
would be in disagreement with experimental upper limits
set by EGRET on \gr emission from nearby clusters of galaxies. 

\item Based on the above
considerations, we set an upper limit on the efficiency of 
conversion of shock ram pressure into energy of relativistic
CR electrons of order 
$\sim 1 \times [\log(p_{max}/m_ec)/\log(4\times 10^7)] ~\%$,
where $p_{max}$ is the maximum energy of the accelerated electrons.

\item IC \gr emission from clusters of galaxies could be 
of the same order of magnitude as \gr emission from $\pi^0$-decay.

\end{itemize}

\section*{Acknowledgments}
This work was carried out at the Max Planck Institut 
f\"ur Astrophysik under the auspices of 
the European Commission for the `Physics of the Intergalactic Medium.
I am indebted to O. Reimer for allowing me to use his unpublished 
results. I am grateful to A. Strong, O. Reimer, T. En{\ss}lin and
T. W. Jones for useful comments on the paper.
The Max-Planck-Gesselschaft Rechenzentrum in Garching is acknowledged 
for precious support. 

\bibliographystyle{apj}
\bibliography{papers,books,proceed}

 \label{lastpage}
\end{document}